# Solidarity should be a core ethical principle of Artificial Intelligence

Miguel Luengo-Oroz[1]
United Nations Global Pulse

*Abstract: Solidarity is one of the fundamental values at the heart of the construction of peaceful societies and present in more than one third of world's constitutions. Still, solidarity is almost never included as a principle in ethical guidelines for the development of AI. Solidarity as an AI principle (1) shares the prosperity created by AI, implementing mechanisms to redistribute the augmentation of productivity for all; and shares the burdens, making sure that AI does not increase inequality and no human is left behind. Solidarity as an AI principle (2) assesses the long term implications before developing and deploying AI systems so no groups of humans become irrelevant because of AI systems. Considering solidarity as a core principle for AI development will provide not just an human-centric but a more humanity-centric approach to AI.*

Over the last few years, dozens of organizations including governments, technology companies, and think-tanks have or are in the process of releasing their ethical guidelines and principles in the development of artificial intelligence (AI) [1]. There is a consensus that AI is a dual purpose technology and we must weigh potential risks and benefits. This is critical, since, hyped or not, we have already witnessed cases where not taking into account minimal ethical considerations in the design and deployment of these technologies have generated civil rights violations in the areas of voting, job, housing or law enforcement, systems prone to discrimination based on gender or race, advancement of autonomous lethal weapons and other "evil" purposes [2]. Certainly, as we continue to use neural network methods which learn by our examples, we will face multiple dilemmas based on principles and values.

It is remarkable that most of the emphasis of these AI guidelines proposals so far is made on developing trustworthy AI - from AI principles approved by the OECD [3] to principles from companies such as IBM [4]. Ethical principles are presented as enablers and core building blocks for a trustworthy AI, that is, they look for AI that works as it should, in many times requiring fairness, explainability, accountability and human agency. Still, AI can be trustworthy but amplify and scale questionable decisions we are taking as humans. Remember that trustworthy nuclear technology requires a non-proliferation treaty for our survival, or trustworthy genetic edition requires a humanity level decision around the modification of human genomes to avoid a biological collapse. AI can be trustworthy and target a weapon that produces civil causalities, AI can be trustworthy and predict a deadly cancer- prediction which might be used by a health insurance not to insure you, AI can be trustworthy and automate the job that millions of people are doing, AI can be trustworthy, know ourselves better than us and operate in the delicate line between suggesting and manipulating- for example- our voting intentions.

---

[1] email: miguel@unglobalpulse.org; address: United Nations Global Pulse, 370 Lexington Ave, New York, NY, US, 10017

[2] The first version of this manuscript was written in February 2019.

Even if the AI principles include do good, be fair, be explainable, be accountable, respect privacy and preserve human agency as proposed recently by the EU [5] – probably the most humanistic approach so far-, they might not mitigate a longer-term risk. As the writer Yuval Noah Harari points out [6], the main struggle for humans in the 21st century might be irrelevance. AI will dramatically augment productivity, and, as he explains (and even if we follow the previous principles), just as mass industrialization created the working class, the AI revolution might create a new irrelevant and unworking class. Enough data capturing any human action will allow making an AI model able to replicate and automate the action almost for free and without need of humans in the loop. In a platform economy where winner takes all, this might be too risky.

Solidarity should be a core ethical principle of the development of AI. Solidarity is understood as sharing the prosperity and the burdens equally, and justly among humans. Solidarity is one of the fundamental values at the heart of construction of peaceful societies and present in more than 30% of world's constitutions [7] and the foundational principles of institutions as the European Charter [8]. It is surprising that while many efforts advocate for an human-centric approach to AI, very few (or none) of the existing ethical guidelines and principles for the development of AI have considered solidarity upfront in this sense. According to a recent review, just 6 out of 84 AI principles guidelines mention the concept solidarity [1]. Furthermore, their definitions are vague and understanding of the concept differs- for example For example the Montreal Declaration [9] proposes the development of autonomous intelligent systems compatible with maintaining the bonds of solidarity among people.

Solidarity as an AI principle (1) shares the prosperity created by AI, implementing mechanisms to redistribute the augmentation of productivity for all, and shares the burdens, making sure that AI does not increase inequality and nobody is left behind; and (2) proposes to assess the long term implications before developing and deploying AI systems so no groups of humans become irrelevant because of AI systems.

From a financial angle, solidarity as a principle will pay back those humans whose data was used to train AI models that learn by the example. A conceptual example can be a royalties system where humans receive a compensation each time an AI system that was trained with their data or with their actions "plays". For instance, human doctors teaching an AI model to diagnose a disease will receive something back each time the model is used to diagnose, or humans writing text pieces to feed an AI automatic speech generator will get something back each time the robot writes something at their style [10]. At a public scale, tax on robots or automation are also options for financial solidarity.

Sharing prosperity implies creating and releasing digital public goods in the form AI models and open sourcing algorithms – from climate models for smart agriculture, to pandemics management. The global AI research agenda needs a compass to direct efforts towards a shared prosperity and achievement of the SDGs, and not to a narrow optimization of internet business or academic benchmarks where most of AI talent is now working on. Digital cooperation [11] is needed to also share the burdens. For instance, international regulations at the domain specific level should incorporate the commitment to notify bias, attacks and unexpected functioning states in AI models used at global scale. Similar to the mechanisms we have for declaring a global health emergency, or to conduct investigations of human rights violations, we may now need to prepare ourselves for the declaration of global AI emergencies where the international community will offer its support to address an AI crisis: for example when thousands of deep fake videos with ethnic violence circulate on a day of elections in a country with history of genocide.

Solidarity in the long term thinking implies conducting a risks and harms assessment before embarking in new AI deployments. A clear example of the lack of long term thinking of current state of the AI is the lack of understanding of the climate impact of the computing resources used to train AI models- is it reasonable to throw tons of $CO_2$ emissions to teach a machine to discern photos of cats and dogs in the internet? Shouldn't we establish sustainability policies by which the expected benefits of the AI model should at least outpace its carbon footprint? [12]. Furthermore, thinking long term implies developing international policy instruments such as extending human rights so they include its digital dimension, agreeing on bans for lethal autonomous weapons, working on regulation of global social media companies, or even preemptive bans until societal impacts and regulatory needs are clear - as could be the case for facial recognition technologies [13].

All in all, one of the biggest long term challenges of AI will be how to redistribute the augmentation of productivity so none is left behind and irrelevant. Solidarity as an AI principle can provide a framework and a narrative to face this challenge, so we do not create new and exacerbate existing inequalities. We need a global safety net for AI technologies [6]. Considering solidarity as a core principle for AI development will provide not just an human-centric but a more humanity-centric approach to AI.